\documentclass[useAMS,usenatbib,usegraphicx]{mn2e}

\title[The Z And spin period ]{Activity cycle of the giant star of Z Andromedae and its spin period }

\author[Elia M. Leibowitz and Liliana Formiggini]
{Elia M. Leibowitz$^{1}$\thanks{E-mail:elia@wise.tau.ac.il}
and Liliana Formiggini$^{1}$,$^{2}$ \thanks{E-mail:
lili@wise.tau.ac.il}\\
$^{1}$The Wise Observatory and the School of Physics and Astronomy, Raymond
and Beverly Sackler Faculty of Exact Sciences \\ Tel Aviv University, Tel
Aviv 69978, Israel\\
$^{2}$INAF - Istiuto di Radioastronomia,
Via Gobetti 101, 40129 Bologna, Italy}

\begin{document}
\date{Accepted 2007 December 12. Received 2007 December 11; in original form 2007 December 04}
\pagerange{\pageref{firstpage}--\pageref{lastpage}} 
\pubyear{2007}
\maketitle

\label{firstpage}

\begin{abstract}
We have reanalyzed the long-term optical light curve (LC) of the symbiotic star Z Andromedae, 
covering 112--yr of mostly visual observations. Two strictly periodic and one quasi-periodic 
cycles can be identified in this LC. A P1=7550 d quasi periodicity characterizes the repetition 
time of the outburst episodes of this symbiotic star. Six such events have been recorded so far. 
During quiescence states of the system, i.e. in time intervals between outbursts, the LC is clearly 
modulated by a stable coherent period of P2=759.1 d. This is the well known orbital period of the 
Z And binary system that have been measured also spectroscopically. A third coherent period of 
P3=658.4 d is modulating the intense fluctuations in the optical brightness of the system during 
outbursts. We attribute the trigger of the outbursts phenomenon and the clock that drives it, 
to a solar type magnetic dynamo cycle that operates in the convection and the outer layers 
of the giant star of the system. 
We suggest that the intense surface activity of the giant star during maximum phases of its 
magnetic cycle is especially enhanced in one or two antipode regions, fixed in the atmosphere 
of the star and rotating with it. Such spots could be active regions around the North and the 
South poles of a general magnetic dipole field of the star. The P3 periodicity is half the 
beat of the binary orbital period of the system and the spin period of the giant. The latter 
is then either 482 or 1790 d. If only one pole is active on the surface of the giant, P3 
is the beat period itself, and the spin period is 352 d. It could also be 5000 d if the 
giant is rotating in retrograde direction. 
We briefly compare these findings in the LC of Z And to similar 
modulations that were identified in the LC of two other prototype symbiotics, BF Cyg and YY Her.  
\end{abstract}

\section{ Introduction}
Some thirteen years ago we made a first attempt to extract information about the
prototype symbiotic star Z And by applying techniques of time-series analysis 
to its optical light curve (LC) (Formiggini \& Leibowitz (1994), hereafter paper I). 
We have collected all historical and modern measurements of the brightness of the 
star known to us, mostly visual estimates by countless amateur astronomers all over
the world. In this we have been greatly helped by the rich and invaluable
data archive of the American Association of Variable Star Observers
(AAVSO). The LC analyzed in that work covered 98--yr of the star history.
Since our first analysis, data on additional 14--yr of this star
photometric behavior have been accumulated. This fact, together with some
experience and knowledge that we gained on the photometric variability of
two other prototype symbiotics, BF Cyg and YY Her,
(Leibowitz \& Formiggini 2006--hereafter paper II; Formiggini \& Leibowitz 
2006--hereafter paper III), prompted us to make a second attempt to understand  
Z And through its long-term LC. In this work we present the results of our 
reanalysis of the more extended LC, now covering 112--yr. Some of our 
conclusions reconfirm part of what we asserted in paper I. Here we present also 
new conclusions about the system.

\section {The symbiotic system Z And}

The basic parameters of the Z And symbiotic star and the common knowledge 
and understanding of this system are summarized in many papers (e.g. Kenyon
\& Webbink 1984; Mikolajewska \& Kenyon 1996; Sokoloski et al. 2006).
Important developments in the study of
this system in the thirteen years since the publication of paper I, that are
related to our present analysis, may be summarized as follows:
From polarimetric data, Schmid \& Schild (1997) were able to determine the 
orbital plane inclination of the binary system {\it i}=47$^{\circ}${$\pm$}12$^{\circ}$. 
The spectroscopic orbital period  was found to be identical to the photometric 
one derived in paper I,(Mikolajewska \& Kenyon 1996; Fekel et al. 2000). 
Sokoloski \& Bildstein (1999) discovered coherent 28 minute oscillations, 
which were interpreted as reflecting the spin period of the  magnetic white 
dwarf (WD) component of the system. This episode was detected, however, only 
during a decline from one upward fluctuation in the brightness of the star 
(within our H6 event--see below) that started in 1997. Another attempt to 
observe rapid oscillations during the 2000-2002 fluctuation of that event 
yielded a null result (Sokoloski et al. 2006; Gromadzki et al. 2006). 
Recently, Sokoloski et al. (2006) made a thorough analysis of multi-wavelength 
observations of the 1996-97 and the 2000-02 fluctuations of event H6. 
They propose that the first one is probably a cataclysmic instability 
episode in an accretion disk around the WD of the system. They suggest that 
the second, much more intense outburst, and presumably also many other similar 
fluctuations of Z And, constitute a new type of a "combination nova" outburst. 
This is an eruption and a large pour of energy from the system, the origin of
which is disk instability, coinciding or preceding an enhanced nuclear burning 
on the surface of the WD.

\section {The long-term light curve of Z And: the data}

For the construction of the visual LC of Z And, we used the Payne-Gaposhkin (1946) 
data, covering the period from 1895 to 1944  and  the visual magnitude estimates 
of AAVSO. Our procedure in bringing the two data sources 
to a satisfactory common scale and zero level is described in paper I--Section 2.

\begin{figure*}
\includegraphics[width=180mm]{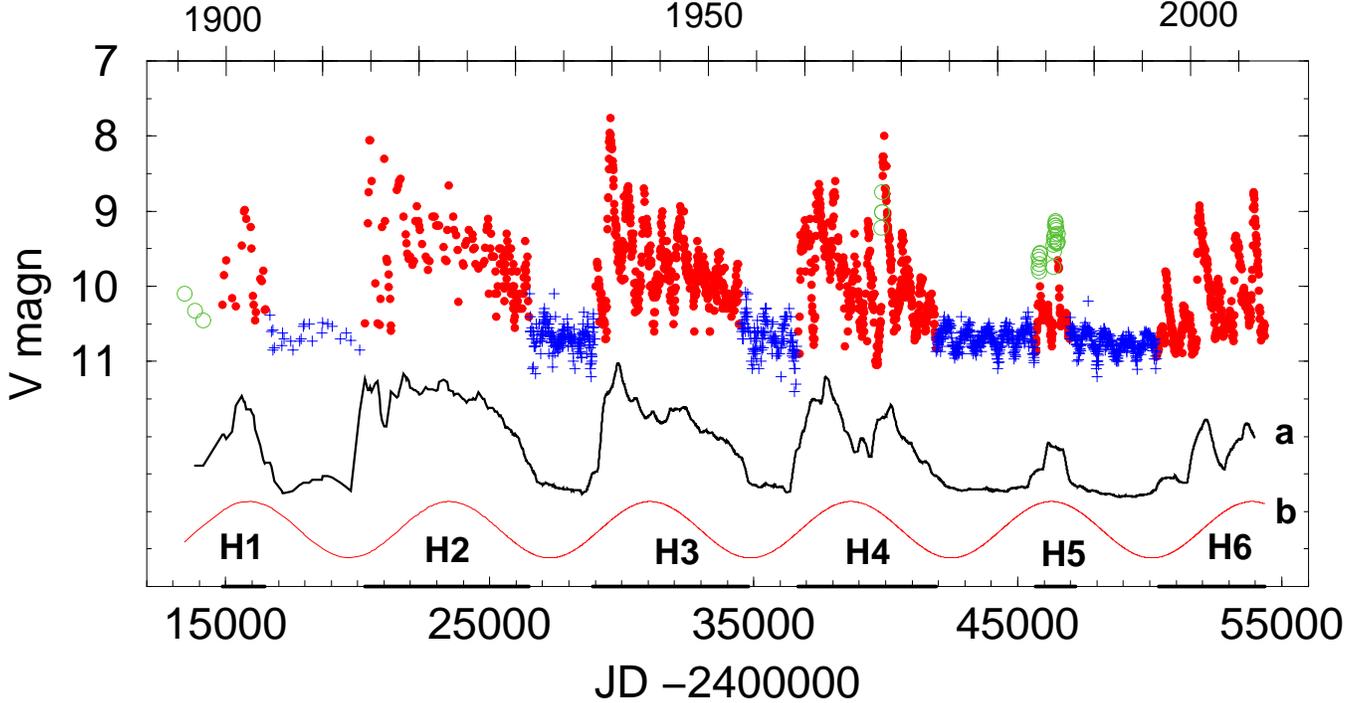}
\caption{Upper curve is a visual light curve of Z AND, from the year 1895 
up until Sept 2007. Crosses are measurements in the low (L) state of 
the system. Dots are the high (H) state. Circles are outlier points 
discussed in Section 4.4. The curve (a) is the running mean of the curve 
above it with a 760 d wide window. The curve (b) is a sine wave of the 
{\em P}1=7550 d period. The two lower curves are shifted down to facilitate 
eye comparison. See text for further details.}
\end{figure*}

The LC shown in Fig. 1 here is the same one presented in fig. 1 of paper I, 
extended up to the month of September 2007. The points displayed in the 
figure after JD=2425000 are the magnitude estimate values in equal time 
bins of 10 d width. Due to the scarcity of the earlier data, those points 
are presented as given in the original sources. In our analysis of the data 
we have eliminated the first three points marked in the plot by open circles, 
on account of their large separation in time from the rest of the data. 
Our results, however, do not change significantly if we add these points to 
the LC. The other twenty nine points marked by circles will be discussed in  
Section 4.4. The curve underneath the observed data points displays the 
running mean of the LC above it. The running window that we have used is 
760 d wide. The sine wave curve depicted in the figure will be explained 
in Section 4.1. The two lower curves were shifted downward from their 
nominal y values to facilitate eye comparison. 

\section {Analysis}
Basic characteristics of the optical LC of Z And become qualitatively
apparent by just looking at Fig. 1. The temporal behavior of the star is
marked by six outburst events, during most of which the mean brightness 
of the system rose by more than one magnitude. This is clearly seen in 
the running mean curve in Fig 1. This curve displays the mean over 
760 d of the star magnitude, namely, the average magnitude of the 
star over one binary orbital cycle of the system (see below). 
Superposed on these brightening episodes are large fluctuations that 
reach up to two magnitudes peak to peak amplitude. The outburst episodes 
which we designate as H1 to H6, are marked by the heavy lines along the 
time axis.

The individual events (red dots in Fig.1) are however quite different 
from each other, in their duration, in their structure, as well as in 
the amplitude of the fluctuations. We refer to all of them as the high (H) 
states of the system. 

In between the H states there are clear time intervals during which the
brightness of  the system is down and in a much quieter state. These are
the low (L) states  of the system (blue crosses in in Fig.1).
 
It can clearly be seen in Fig. 1 that the brightness of the star at all L 
states is cyclically modulated with one and the same periodicity.

\subsection {Outbursts}

Our first quantitative treatment of the LC was a computation of its power
spectrum (PS) (Scargle, 1982). We also applied on it the AoV technique of
period search in time-series (Schwarzenberg-Czerny 1989). Both yield very 
similar results.

\begin{figure}
\includegraphics[width=90mm]{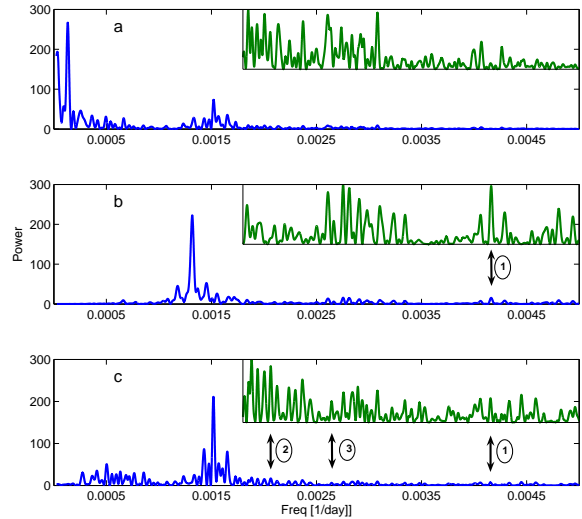}
\caption{ Power spectra of the light curve of Z And.
(a) Observed light curve shown in Fig. 1; (b) Low (L) state; 
(c) High (H) state; Inserts are blow-ups in the y direction of the section 
of the power spectrum below them. Arrows and further details are explained 
in section 6.2.1.}
\end{figure}

Fig. 2(a) presents the PS of the observed LC seen in Fig. 1, in the period 
range from 200--d to the entire length of the LC, nearly $\sim$40000 d. 
There is a high rise of the spectrum at the extreme low frequency region of it. 
It indicates some variability on time scale of the length of the LC. 
This can either be a real effect or an artifact due to some systematic error 
in our procedure of normalization of the different data sources in bringing 
them to a common magnitude level.

Asides from this feature, the PS  is dominated by two peaks, both are 
statistically significant. The major one is at the frequency f1,
corresponding to the period  {\em P}1=7550$\pm$722 d. The uncertainty quoted  here
and throughout  this work is P$^2$/T,  P  being the corresponding period 
and T is the time length of the LC.
The second high peak in the PS near f=0.0015 d$^{-1}$ will be discussed in Section 4.3.

The lowest curve in Fig. 1 is a sine wave of the {\em P}1 periodicity,
fitted to the data by least squares. The purpose of presenting it is only to show 
qualitatively how faithful is {\em P}1 in representing the quasi cycle of the 
outburst occurrences.

We note that our present {\em P}1 value and the value quoted in paper I for the
characteristic repetition time of the outbursts of Z And are within the
corresponding error estimates. The difference between the two values is due
to the new outburst of the system of the present era, that was added in the
last fourteen years to the older LC analyzed in paper I. In any event, as we
shall discuss in Section 5.2,  we do not claim, nor do we believe that the
{\em P}1 number represents a period of a strictly cyclic process. The outbursts
of Z  And take place quasi-periodically and {\em P}1 is presently the best
estimate for the average time interval between centres of successive
outburst events.
The inserts in Fig. 2 (a)--(c)
are blowups of the high frequency end of the corresponding PS. They, and the 
arrows in all frames of Fig. 2 will be discussed in Section 6.2.1.

\subsection {The L state}
Fig. 2(b) is the PS of the combined LC of all low states of the system. The
unquestionable highly statistical significant peak is at the frequency f2,
corresponding to the period {\em P}2=759.1$\pm$8.6 d. Fig. 3(a) displays twice the L
state LC folded onto the {\em P}2 periodicity. It is quite evident that the
system has preserved the phase of the {\em P}2 cycle, as well as the amplitude of
this variation, throughout the entire 112--yr of the observations, i.e.
for more than 44 successive cycles. This is in spite of the six time
intervals of different lengths, those of the H state of the system, that
are dispersed within these 112--yr, during which the P2 periodicity all
but disappeared from the LC of the star. Note the absence of any spectral 
feature in Fig 2(c) at the frequency of the high peak seen in Fig 2(b) (see Section 4.3) 

\subsection {The H state}
Fig. 2(c) is the PS of the H state LC. Before combining the magnitudes of
all the outbursts of the star into this single H LC, we subtracted from
each individual H LC a polynomial of first or second degree.  This process
detrended each individual H LC from variability on the time scale of the H
time interval itself. When we subtract from each H LC only the
corresponding mean magnitude we obtain similar results.

The highest peak in Fig. 2(c), which is statistically highly significant, is
at the frequency f3, corresponding to the period {\em P}3=658.4$\pm$5.5 d. 
The cluster of lower peaks around it are either aliases of the high one or 
resulting from the highly non harmonic structure of the {\em P}3 periodicity. 
For example, the second highest peak to the right of the f3 one, corresponds 
to the period of 606 d. This is the beat period of {\em P}3 and {\em P}1, due 
to the gaps in the H LC, (the time intervals of the L state), that break up the 
H LC with the 7550 d periodicity.

Fig. 3(b) is the LC of the H state, folded onto the {\em P}3 periodicity. The 
twenty nine points marked by  red circles are the ones marked by circles in Fig. 1. 
We shall refer to them shortly. With the exception of these points, all other 1533 
points seen in the figure are evidence of the persistence of the {\em P}3 period 
in the H state of the system. The figure demonstrates that the period {\em P}3 has 
the same phase in all its appearances, including the H6 event and the H1 event. 
These two events are separated from each other by 55 cycles of this period and 
the $\sim$100 years that separate them contain the five time intervals of the 
L state, during which the {\em P}3 period is absent from the LC. There is no 
peak at the f3 frequency in the PS of the L state. Note the saw tooth structure 
of the mean profile of the {\em P}3 photometric cycle. 

\begin{figure}
\includegraphics[width=90mm]{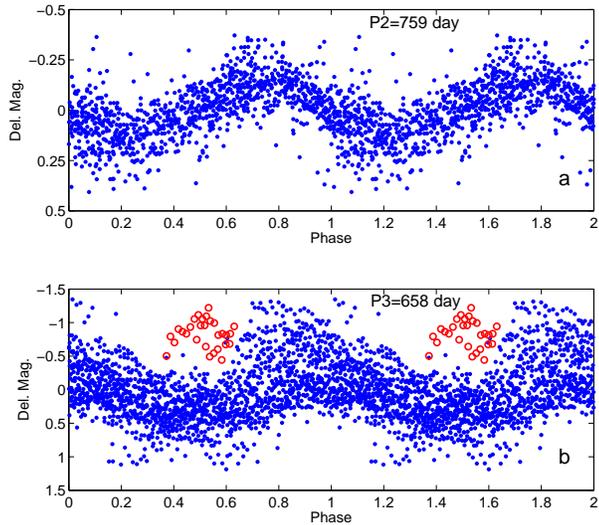}
\caption{ (a) Light curve of low (L) state folded onto {\em P}2= 759 d, showing 
the cycle twice. (b) Light curve of high (H) state folded onto {\em P}3= 658 d. 
The open circles mark the outliner points (see section 4.4).}
\end{figure}

Fig. 4 displays in details the quality of the fit of the P3 periodicity to the 
individual fluctuations in each of the six outburst events. Each frame presents 
the observed detrended LC of one event. The harmonic function is a sine wave 
fitted to the combined H LC of the star. Thus the amplitude and the phase of 
the wavefronts are the same in all six frames.  Note that the scale of the x 
and y axis in Fig. 4 (e) is different from all the rest. Fig. 4 (e) presents 
a time range that is wider than event H5, as defined in Fig. 1. The outburst 
event itself is the segment delineated by the two vertical lines in the figure. 
The dotted line is the run of harmonic wave of the {\em P}2 periodicity that 
is fitted by least squares to the L light curve. 

In all other frames, i.e. all H events except H5, one can see in the figure 
that while the structure of the observed individual fluctuations varies 
considerably and is markedly different from the smooth sine wave structure, 
the phase seems to fit very well. Four points in H4 (surrounded by green ellipse in Fig. 
4 (d)) around 
JD 40000 are exceptions, showing a sharp rise in the brightness of the star 
that starts near a minimum of the {\em P}3 periodicity. We shall return to 
these four outliner points, as well as to Fig. 4 (e) in the next section. 

\begin{figure}
\includegraphics[width=100mm]{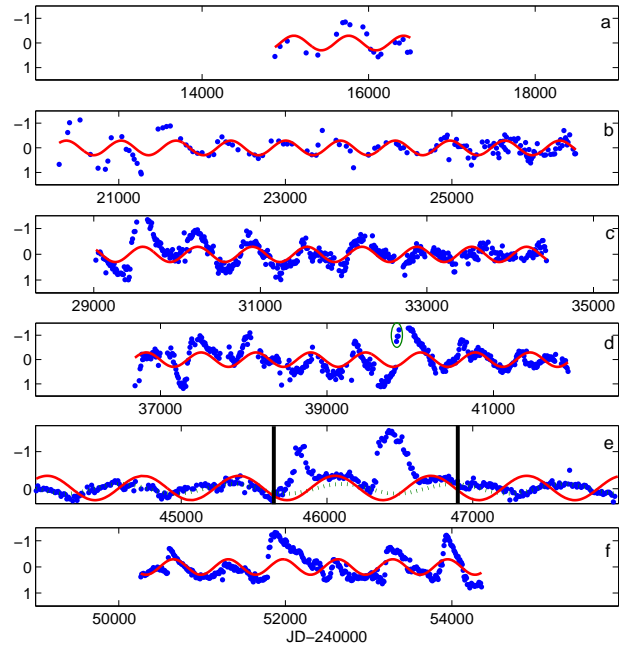}
\caption{The six high state events of Z AND. The points are the detrended LC, 
the solid line is an harmonic wave of the period {\em P}3 = 658 d, best-fitted 
to the observed points. Note the different scale of frame (e), and see 
section 4.4 for further details. Green ellipse in frame (d) circulates the four 
outlying points of event H4 referred to in the text.}
\end{figure}

\subsection{The outlier points}

Most of the 29 H state points that are out of phase with the 1533 measurements 
that exhibit the {\em P}3=658 d oscillations, are those that constitute the two 
peaks of event H5. They are shown in details in Fig. 4 (e). The figure presents 
a wider segment of the LC. The H5 event as we identify it, is delineated by the 
two vertical lines in the figure. The solid sinusoid line is the harmonic wave 
of the {\em P}3 periodicity that is fitted by least squares to the entire H LC 
of the star, as in all other frames of Fig. 4. The dotted line is the 
{\em P}2=759 d harmonic wave fitted to the overall L LC. One can see in Fig. 4(e), 
that outside the two vertical lines the observed points follow very accurately 
the dotted line. In fact most of the dotted line is obscured by the observed 
points that in the resolution of the figure fall right on top of it. The LC 
in the section between the lines is composed of two well defined peaks and 
a sequence of lower points in between them, and a few that follow the second 
one. These lower points too exhibit two oscillations, the maxima of which are
 systematically higher by 0.1-0.2 magnitude than the maxima of the 759 wave. 
They also follow much better the solid line representing the 658 wave. 
Thus, the underlying rise in the brightness of the star of the H5 event, 
as compared with the brightness level in the L state, seems to follow 
the 658 periodicity, in phase with all other H events. The two peaks of this 
event, as well as the four outlier points in event H4, seem to be some 
sporadic flaring activity of the system that is probably related to the 
magnetic active phase of the star (see section 5.3) but occur not in phase 
with the 658 d periodicity. 

For the first fluctuation of the H5 event, the exceptional nature of it 
is apparent even in the photometric data. Fig. 4 (e) shows that this peak 
is different from all others. Its duration is only one half of the 658 d
 period which is the typical duration of nearly all other fluctuations 
observed in the H state of the system. 

The two out of phase fluctuations of H5 will be discussed further in Section 6.1.

\section{Discussion}

\subsection{The low state}

The optical brightness of Z And in its low state is clearly
modulated by the {\em P}2=759 d periodicity. This period coincides
well with the spectroscopic periodicity found for this
binary star (Mikolajewska \& Kenyon 1996; Fekel et al. 2000) and it is therefore 
well established as the orbital period of the system (paper I).  The profile 
of the optical photometric cycle is nearly sinusoidal, with minimum light 
obtained near spectroscopic conjunction, when the giant star is in front of 
the WD of the system (Sokoloski et al. 2006). The cool component at this phase 
is obscuring part of the emission nebula, the excitation centre of which 
is the WD. This binary  configuration is also the phase of minimum light 
reflection from the hemisphere of the giant illuminated by the hot 
component (Formiggini \& Leibowitz 1990). 

\subsection{The outburst cycle}
Six outbursts of Z And have been recorded so far. They are
quite different from each other in duration, in the amount
of excess optical radiation emitted in them, as well as in
their structure. The outbursts seem to take place, however,
with a fairly constant time interval of some 7550 d, i.e.
some twenty one Earthly years, between the centres of successive
ones. An outburst is generally manifested photometrically in
the optical region by one or more magnitude increase in the
average brightness of the system, and by intense periodic
oscillations with a period {\em P}3=658.4 d of strongly varying
amplitude.

Very similar features characterize the outbursts of two
other symbiotic stars, BF Cyg (paper II) and YY Her (paper III).
In these two systems the triggering of the outbursts has been ascribed to
magnetic activity in the outer layers of the giant component
of these two binary systems. A solar-like magnetic dynamo
mechanism has been suggested as the driver of the phenomenon
and as the clock that maintains the quasi periodic cyclic
nature of it.  The similarities in the
phenomenology suggest also a similarity in mechanism. We
propose that in Z And too, each of the six recorded outbursts
resulted from intense, solar-like magnetic activity that
takes place quasi-cyclically in the giant star of this
system.

The quasi-periodic nature of the magnetic cycle in the giant
of Z And is not dissimilar to the timing of the magnetic
activity of our own Sun. The solar cycle is also far from
being strictly periodic, with intervals between successive
maxima varying between nine  and thirteen years (Babcock 1961;
Fligge, Solanki \& Beer 1999; Mursula \& Ulich 1998). Similarly, vast
variations in the format of different maxima of the activity
cycle as observed in Z And, characterize also the activity
cycles of our Sun (Solanki et al. 2002).

\subsection{The High state}

During outbursts, the brightness of the star fluctuates
violently with a distinct periodicity {\em P}3=658.4 d. While
the structure of these oscillations is far from being
stable, the star seems to preserve the periodicity and the
phase of their occurrences. With the two exceptions
discussed in Section 6.1, the clock that controls the {\em P}3
cycle appears to be stable throughout the $\sim$100 years period
of the observations, i.e. for more than 55 cycles of the 658 d periodicity.  

The persistence of the {\em P}3 periodicity and phasing, implies
that the clock controlling them must be a rather stable one.
In addition to the binary period, another clock in the system 
that is able to withstand the
violent activity of the system with the tenfold or more
variation in its output power, could be the spin of the giant
star. We suggest that the period of the observed outburst
oscillations and its stable phase are the combined effect of
these two coherent cycles of the system, according to
the following qualitative model.

\subsubsection{The qualitative model}
The system consists of a hot white dwarf (WD) star and an M
giant, in binary orbit around each other with the period
{\em P}2=759 d. Sokoloski et al. (2006) showed quite convincingly that
the violent activity of the 2000-2002 Z And fluctuation, and
presumably also of the other violent fluctuations of the system,
consist of a nuclear runaway process (NRA) on the surface of the
WD, combined with instability episodes in an accretion disk
around the WD that seems to precede it or to coincide with it.

We suggest that the cause and the trigger of these two
processes that constitute the "combination nova" events is 
the enhanced activity of the giant star,
regulated semi-periodically by the giant inner magnetic
dynamo. One of the consequences of this activity is an
intensification of the giant stellar wind, as is the case for the Sun.

In the case of our Sun, detailed measurements of solar wind various 
parameters in the last few years, especially by {\it Ulysses}, {\it SOHO} and 
other space probes, reveal this phenomenon as a multi-components, 
complex process. However, a global structure, in mass density as 
well as in velocity, that is regulated to a very large extent by 
a dipole magnetic field that can be ascribed to the Sun, is clearly 
apparent. During maximum activity seasons of the solar cycle, the 
wind pattern is changing, when both density and velocity taking 
different spatial configuration, also governed, however, by a 
global magnetic dipole field. For recent review of solar wind 
dynamics over the solar cycle see Schwenn (2006). Notice in 
particular Fig. 5 of that paper. During maximum activity seasons 
of the solar cycle, for example, the wind velocity from the magnetic
 pole regions is strongly  enhanced (Marsch 2006; Ofman 2006) .

It is very much likely that
for the giant of Z And as well, during active seasons, stellar wind 
from the poles of a dipole magnetic field of the star is
more enhanced, in velocity, in mass density or in both, than the wind 
from the magnetic equatorial regions. 
The same qualitative model would apply if the magnetic field near the poles 
is inhibiting mass loss rather than intensifying it, and the wind from the 
magnetic equator is the more enhanced component. We hypothesize that the 
axis of the dipole is inclined with
respect to the spin axis of the star and rotates with it.
This is the case in the Sun, although the inclination angle there is 
quite small (Babcock 1961; Bilenko 2002). However, the phenomenon of 
magnetic oblique rotators with large inclination angles is not uncommon 
among stars (Kurtz 1990). We should note, however, that the existence of 
the hypothesized magnetic dipole field of the giant star would not be 
easily detected directly. In the solar case, the dipole field that is 
associated and presumably controls the global structure of the wind, 
reaches a maximum intensity of no more than 50 Gauss. The local fields 
in spots and in other features of surface and coronal activities
are orders of magnitude stronger. For a distant observer, as we are with
respect to the giant in Z And, they mask any observable diagnostic of the
global field, such as measurable polarization or the Zeeman effect. The
polarization observed by Schmid and Schild (1997) is due to Raman
scattering and the geometry involved is on a much wider scale of the 
Z And nebula that far exceeds the binary dimensions. We further note, 
however, that the essential requirement here is not the dipole magnetic 
field. What is needed for our interpretation is the existence on the 
surface of the giant of one, or two antipodal active regions, fixed in the 
outer atmosphere and rotating with the star.

F$\acute{e}$rnandez-Castro et al. (1988) showed that even during
quiescence states of Z And, some nuclear burning is taking
place on the surface of the WD. This implies that even at
quiescence, the giant is loosing mass at a large rate. This
would indicate in turn, that the giant star is not far from
filling its gravitational Roche lobe of the binary system 
(Sokoloski et al. 2006).
The gravitational pull of the WD creates a tidal wave in the
outer layers of the giant. The high tides are fixed in the rotating 
frame of the system. If the giant rotation is not locked to the orbital 
cycle, the tidal  wave is traveling in the atmosphere of the
spinning giant. For an observer on the surface of the giant,
the period of one cycle, namely, the synodic period of the wave, is
the beat of the binary orbital period of the system and the sidereal spin 
period of the star (see next Section). 

During phases of maximum activity of the giant, whenever the bulge 
of the tidal wave is sweeping across the neighborhood of either
the North or the South magnetic poles of the star (or, one might say, 
when the magnetic pole is sweeping through the tidal bulge) a
specially intense mass loss rate is ensued. This results in
enhanced cataclysmic dynamical processes in the disk around
the WD. It also increases the rate of supply of fuel for the 
nuclear runaway burning on the surface of the WD, 
and thus greatly intensifying it. 
Due to the strong dependence of the rate of nuclear burning on the surface 
of a WD on temperature and density, even small enhancements in the 
accretion rate onto the disk and eventually to the WD surface are enough to 
excite vast variations in the energy output rate of the system 
(Sokoloski et al. 2006). This enhancement may occur
once every synodic period of the traveling tidal wave, if
the magnetic activity near one pole is significantly more
enhanced than near the other. It may occur twice every
synodic cycle if the two antipodal regions on the surface of
the giant are more similar in their surface activity. This
qualitative picture explains well the phase preservation in
the {\em P}3 periodicity in all outburst events of the system
(with the exception of the flares mentioned above). It is the 
coherence of the binary cycle on one hand, and that of the spin period of the
giant on the other, that ensure it.

\section{The giant spin period}

We observe that during outbursts, the period of the activity
cycle is {\em P}3=658 d. This is, as explained, the synodic
period of the tidal wave in the giant atmosphere.  If there are two 
antipodal centres of activity on the surface of the rotating giant, 
the synodic period is in fact 1317 d. For a binary
orbital period of {\em P}2=759 d, this would be the case if 
the sidereal period of the giant spin is 482 or 1790 d.
If the intense wind at maximum activity season is emanating mainly 
from just one pole, the giant sidereal spin period is 352 d. In this 
case it could also be 5000 d if the giant rotates in the retrograde direction.

\subsection{The H5 maximum activity season}

According to our suggested scenario (Section 5.3.1), during seasons of 
maximum magnetic activity of the giant, whenever a magnetic pole is 
sweeping through the tides bulge, a large flux of matter is flowing 
from the giant onto the Roche lobe of the WD. Thus, during the giant 
activity season, what is being supplied in large quantities to the 
disk around the WD is not only mass but also a great deal of angular 
momentum. The disk instability, which according to the combination 
nova scenario (Sokoloski et al. 2006) precedes the NRA on the surface 
of the WD, may be a response of the disk to the sudden increase in 
the influx of angular momentum, no less than to the increase in the 
rate of mass supply. This may be an additional reason for the prompt 
response of the disk to the changing rate of mass accretion.

In contrast, the two out of phase eruptions of H5 may be understood as 
instability events that erupted in the disk spontaneously. In this event 
too, the supply of matter to the disk was modulated by the P3 periodicity, 
in phase with the oscillations in all other H events, as evident in the 
low amplitude variation of the low lying points of the H5 event, as 
described in Section 4.4. 
However, the amount of material supplied to the disk in this event, 
and hence also of angular momentum, was obviously smaller than in all 
other H events. Due to the smallness of these two parameters, they 
disturbed the disk less sharply than in the common, more powerful maximum 
activity seasons of the giant. By saying that in H5 the disk instabilities 
occurred spontaneously we mean that the conditions for instability developed 
in the disk more slowly. The timing of their onset was regulated by the 
internal physics of the disk itself, as is the case in the common disk 
instability events in dwarf novae. This is why they are not in phase 
with all other P3 oscillations.

The small amplitude brightness variation during H5 that is in phase with 
the P3 periodicity possibly reflects the brightness response of the disk 
to the increase varying supply of material from the giant. It may also be 
the result of modest variations in the nuclear burning rate on the surface 
of the WD, due to variations in the rate of materials falling onto the WD 
surface either directly or after spiraling through the accretion disk. 
 
\subsection{Further possible results}

In this section we report on two other findings in the LC of the star 
that are very much in line with our suggested scenario. The two additional 
features that we identify in the PS are, however, clearly below an appreciable 
statistical significance. We therefore do not consider them as proofs of 
our model. We do believe, however, that it is worthwhile mentioning them 
here for what they are, since if their reality is better confirmed by future 
observations, they do provide considerable support for our interpretation.

\subsubsection{Sidereal spin signal}

The inserts in the three frames of Fig. 2 are blow-ups in the y direction of 
the high frequency ends of the corresponding PSa. The arrows in frames (b) 
and (c) designated "1", are pointing at peaks corresponding to the periods 
240.23{$\pm$}.86 d in the L PS (Fig. 2 (b)), and 240.6{$\pm$}0.73 d in the
 H PS (Fig. 2 (c)). Well within the error in the peak position these number 
are precisely half the 482 value, one of the candidates for the sidereal 
spin period of the giant. A detection of a signal with 241 d periodicity 
is expected if there is a concentration of bright or dark spots around the 
magnetic poles of the rotating giant.
Fig. 2 (a) shows, however, that the 241 d peak is absent from the PS of the 
entire LC, namely, the curve of the combined sequence of the low and the 
high states. This means that if the {\em P}=241 d is a real feature of 
Z And, there is a phase shift in this periodicity between the low and the 
high states of the system. If the 241 d periodicity is indeed half the 
spin period of the giant this would indicate that the hot or dark spots 
on the surface of the giant that are responsible for this light variations,
change their location on the surface of the star between the two states of 
the system. For example they could be concentrated more around the poles in 
one state, moving towards the magnetic equator in the other. We note also 
that in the PS of the high state (Fig. 2 (c)) there is a weak signal also 
at {\em P}=484.3{$\pm$}2.97 d (peak "2").

Due to the low statistical significance of these signals we refrain from 
using them for making a choice between the possible values that we suggest 
for the sidereal spin period of the giant.

\subsubsection{Ellipsoidal effect}

Mikolajewska et al. (2002) report on preliminary results of IR observations 
that indicate the presence of an ellipsoidal effect in the LC of Z And. If 
the star is nearly filling its Roche lobe, a significant distortion of its 
outer layers from spherical symmetry is indeed expected. Such a distortion 
would be reflected in the LC of the system if the contribution of the giant 
photospheric light to the total optical luminosity of the system in the 
observed spectral region is not entirely negligible. We do find in the 
H LC a weak signal of a period which is half the orbital period, as 
expected for the ellipsoidal effect. Arrow "3" in Fig. 2 (c) is pointing 
at a peak in the PS corresponding to the period 378.2{$\pm$}1.8 d. 
The peak is below any statistical significance and it is undetectable 
in Fig. 2 (b), the PS of the L LC. We therefore do not consider it as 
a positive detection of the ellipsoidal effect as stated above, in spite 
of the near perfect coincidence of its frequency with twice the orbital 
frequency. We nonetheless wanted to find out where an assumption that it 
is a true sign of the ellipsoidal shape of the giant could lead us.  

Since the giant spin is not locked to the orbital revolution, the tidal 
bulge is noncollinear with the centres of the two stars. If the spin 
period is 482 or 352 d, the orbital angular velocity is slower than 
that of the spin and the bulge is leading in front this line 
(Lecar, Wheeler \& McKee, 1976). By least squares we computed the sine 
wave of half the orbital period that best-fit the H LC. We compare this 
LC to the sine wave of the orbital period itself that is best fitted to 
the L state of the system. We found that one minimum of the "ellipsoidal" 
LC is preceding the orbital minimum by 130 d. The other "ellipsoidal" 
minimum is obviously preceding the maximum of the orbital cycle by the 
same number of days. In units of angle this means that the direction of 
the high tides protrusion is leading the binary centres line by $\sim$60$^{\circ}$. 
If the reality of the {\em P}2/2 signal could be confirmed, this result 
would be an important source of quantitative information about dissipative
processes in the outer layers of the giant, those that are responsible for 
this noncollinearity.

\section{Comparison with two other symbiotics}

We have performed in the past time-series analysis similar to the one 
presented in this work, on the historical LCs of two other prototype 
symbiotics, BF Cyg and YY Her (paper II and III). The three symbiotic 
stars are very similar in the basic astronomical parameters that 
characterize the corresponding binary systems. Table 1 presents some 
of these parameters for the three stars. Thus Z And seems to become 
a third symbiotic system in which evidence for the operation of a 
solar-like magnetic dynamo cycle has been discovered. The stars 
harboring the process in the symbiotic systems are giants, whereas 
the Sun is a main sequence star. The intensity of the surface activity 
of the symbiotic giants is probably also different from that of the Sun, 
being of a more grandeur scale in the giants. Note, however, that the 
dramatic effects of the activity cycles of these stars are mainly due 
to the fact that they are members in interacting binary systems. 
The activity cycle on the giant is mainly a trigger for the vast 
outpour of energy, which is generated in the extreme conditions on 
and around the WD component of the symbiotic system. The quasi-periodicity 
of the phenomenon in all four stars is however of the same order of 15-20 yr.

The rotation rate of the giant of Z And is much more distant from 
synchronization with the orbital cycle than in the other two symbiotics. 
Thus the synodic period of the tidal wave in its outer atmosphere is much 
shorter than the dynamo magnetic cycle. In BF Cyg and YY Her, the near 
synchronization makes the spin-orbit beat period to be of the same order 
as that of the magnetic dynamo cycle. This is why in Z And we do not 
find in the LC any trace for an effect of the tides on the stellar 
magnetic cycle itself. This is also the case in the Sun with its spin 
period of merely $\sim$27 d. The tides in the Sun, excited primarily 
by Jupiter, are obviously also of a much smaller scale than in the symbiotics. 
In any event, in both Z And and in the Sun, over one activity cycle, 
the displacement vector due to the radial motion induced by tides averages 
to zero, since one activity cycle includes many tidal oscillations. 
In BF Cyg and YY Her, on the other hand, the radial oscillations of the 
outer envelope of the giant that are due to tides have the same characteristic 
time as that of the magnetic cycle itself. The role of tides in the creation 
and maintenance of the dynamo may therefore be comparable to that of 
differential rotation, convection and meridional flow that are currently 
believed to be at the dynamical roots of the magnetic dynamo phenomenon 
(paper III and references therein). This is why in these two stars the 
synodic period of the tide wave is manifested in the timing of the 
magnetic dynamo cycle of their corresponding giant components.

The giant of Z And is thus joining the class of stars that harbor magnetic 
dynamo process, as have been suggested already in the past 
(Soker 2002, Soker \& Kastner 2003). In Z And, as well as in BF Cyg and 
YY Her, the observational effects of this phenomenon are gaining very 
large magnification, due to the binary interaction with the nearby WD 
companion. This strengthens the proposition that symbiotic stars may 
turn out to be important laboratories for studying the stellar magnetic 
dynamo cycle, an astronomical process that affects Earthlings directly.

\section{Summary}

New analysis of an updated optical LC of Z And covering the last 112--yr 
of the star history reestablishes the well known 759.1 d photometric 
periodicity of the star, known to be the binary orbital period of the system. 
The analysis confirms the semi-periodic nature of the outburst phenomenon 
of this symbiotic star. The characteristic time interval between the centres
 of successive outbursts is 7550 d. It is suggested that the quasi-periodicity 
of the outburst events is regulated by a solar-type magnetic dynamo process 
in the giant of the system. Enhanced mass loss during maximum phases of 
this activity cycle is the trigger and provides the fuel for the intense 
energy outpour of energy during outbursts, emanating mainly from the vicinity 
of the WD component. In addition to an increase by one magnitude or more in 
the mean brightness of the system, outbursts are also characterized by 
violent periodic oscillations. Our analysis reveals a persistent periodicity 
of {\em P}3=658 d of these oscillations that preserves its phase throughout 
the entire 112--yr of the observations. The oscillations result from periodic 
modulation of the mass accretion rate onto the WD component of the system, 
from the bulge of a tidal wave in the atmosphere of the giant which is fixed 
in the binary rotation frame.  The synodic period of the tidal wave that 
circulates in the atmosphere of the giant is the measured {\em P}3 value, 
or twice this number. This is the beat period of the giant spin and the 
orbital periods of the system. The sidereal spin period of the giant is 
either 482 or 1790 d. It could also be 352 d or possibly  5000 d, 
if the giant is spinning in the retrograde direction.

\begin{table*}
\begin{tabular}{@{}lrrrcrc@{}} \\
\multicolumn {6} {|c|}{\bf Table 1}{Comparison of the properties of Z  And, BF Cyg
and YY Her}\\
\hline
  &  Z And & Ref.  & BF Cyg &Ref.  & YY Her & Ref.   \\

\hline
\\
Giant Sp. Type          &  M4 III      & 1    &  M5 III      & 1    & M4 III   & 1 \\
Luminosity L $\odot$    &   620-1600   & 2    &  5200        & 2    &   1100   & 2 \\
Binary period           &   759.0      & 5    &  757.3       & 3    &   593.2  & 4 \\
Giant Spin period       &   482$^{*}$  & 6    &  798.8       & 3    &   551.4  & 4 \\
Tidal wave period       &   1317       & 6    & 14580        & 3    &  -7825   & 4 \\
Solar-type period       &   7550       & 6    &  5375        & 3    &  4650    & 4 \\
\\
\hline
\end{tabular}\\

*) This is a slightly preferred number among four possible values.

1) M\H{u}rset \& Schmid (1999), 2) M\H{u}rset et al. (1991), 3) Leibowitz \& Formiggini (2006),\\
 4) Formiggini \& Leibowitz (2006), 5) Fekel et al. (2000), 6) This paper.

\end{table*}

\section*{Acknowledgments}

       We acknowledge with thanks the variable star observations from the AAVSO
International Database contributed by observers worldwide and used in this
research. This research is supported by ISF - Israel Science Foundation of the
Israeli Academy of Sciences.

\label{lastpage}

\end{document}